\documentclass[letterpaper, 10pt, conference, leqno]{ieeeconf}
\IEEEoverridecommandlockouts
\usepackage[english]{babel}
\usepackage{upgreek}
\usepackage{amsfonts}
\usepackage{multicol}
\usepackage{amssymb}
\usepackage{amsmath}
\usepackage{cite}
\usepackage{graphicx}
\usepackage{hyperref}

\let\labelindent\relax
\usepackage{enumitem}

\usepackage{titlesec} 
\titlespacing*{\section}
{0pt}{1pt}{1pt}
\titlespacing*{\subsection}
{0pt}{1pt}{1pt}

\titleformat{\paragraph}[runin]
  {\normalfont\normalsize\bfseries}  
  {}  
  {0pt}  
  {}  

\titlespacing*{\paragraph}
  {0pt}{0pt}{1em}  

\titleformat{\subparagraph}[runin]
  {\normalfont\normalsize\bfseries}  
  {}  
  {0pt}  
  {}  

\titlespacing*{\subparagraph}
  {0pt}{0pt}{1em}  

\expandafter\def\expandafter\normalsize\expandafter{%
    \normalsize%
    \setlength\abovedisplayskip{0.15cm}%
    \setlength\belowdisplayskip{0.05cm}%
    \setlength\abovedisplayshortskip{0.15cm}%
    \setlength\belowdisplayshortskip{0.15cm}%
}

\usepackage{xcolor}
\newcommand{\Red}[1]{\textcolor{black}{{#1}}}

\title{\LARGE \bf Two-Layer Attention Optimization for Bimanual Coordination}

\author{Justin Ting, Jing Shuang (Lisa) Li
\thanks{This work is funded by the NSF GRFP. J.T. and J.S.L. are with the Division of Electrical and Computer Engineering at the University of Michigan, Ann Arbor. {\tt\small sigfyg@umich.edu, jslisali@umich.edu}.}
}

\begin{document}

\maketitle

\begin{abstract}
Bimanual tasks performed by human agents present unique optimal control considerations compared to cyberphysical agents. These considerations include minimizing attention, distributing that attention across two isolated hands, and coordinating the two hands to reach a broader goal. In this work, we propose a two-layer optimization problem that explicitly quantifies these considerations. The upper layer solves an attention distribution problem, while the two lower layer controllers (one per hand) tracks a trajectory using the upper layer solution. We introduce a formulation for attention control, where attention is a vector that is bound within a hyperbolic feasible region, determined by specifications of the lower layer controllers. We use this two-layer controller to optimize a single-player game of pong, rallying the ball between two paddles for as long as possible. We find several emergent behaviors from this optimization in simulations; stronger coordination leads to lower long-term attention, while high asymmetry and aggressive centering for stability increase overall attention. 

\end{abstract} 

\section{Introduction}
Bimanual tasks are motor coordination problems that require both hands. In previous works, bimanual control problems (e.g. juggling, devil-sticking \cite{devilstick_control}\cite{hybrid_juggling}) exploit the left-right symmetry of both the dynamics and controller, splitting a two-handed problem into one-handed problems. These problems can be formulated as a hybrid control problem to model the discontinuous collision dynamics. This formulation is useful for implementing the control policies in robots. 

Bimanual tasks are also useful for studying human sensorimotor control policies, which present considerations distinct from cyberphysical systems. In addition to motor control, the brain must strategically send signals to separate limbs \Red{to achieve a broader coordination task.} Furthermore, the brain allocates resource-limited time-varying attention to active sensing and limb actuation, whereas control policies for robots or cyberphysical systems give uniform attention across all sensors, actuators, and time. Optimal control has been used to study human bimanual coordination. As demonstrated in participant experiments \cite{linear_bimanual_control}\cite{sherwoodDividedAttentionBimanual2001}, changing the goals and constraints of a task can change the attention allocated to each arm. The authors of \cite{linear_bimanual_control} could correlate distinct attention policies to LQR gains. In our work, rather than implicit attention in the control policy, we explore strategies that emerge when attention is explicitly quantified. Other works use data from human driving experiments to show how active sensing policies can be described with Inverse Optimal Control (IOC) \cite{maxentropy_IOC_1}\cite{SERD_IOC_driving}. This suggests \Red{that there is a cost to processing sensory data that does not contribute to the success of the task}, and the brain is executing some strategy to minimize this cost. While these works represent the attention strategy as a sensor model, our attention strategy takes the form of modulating cost parameters. 
\begin{figure}\includegraphics[width=\linewidth]{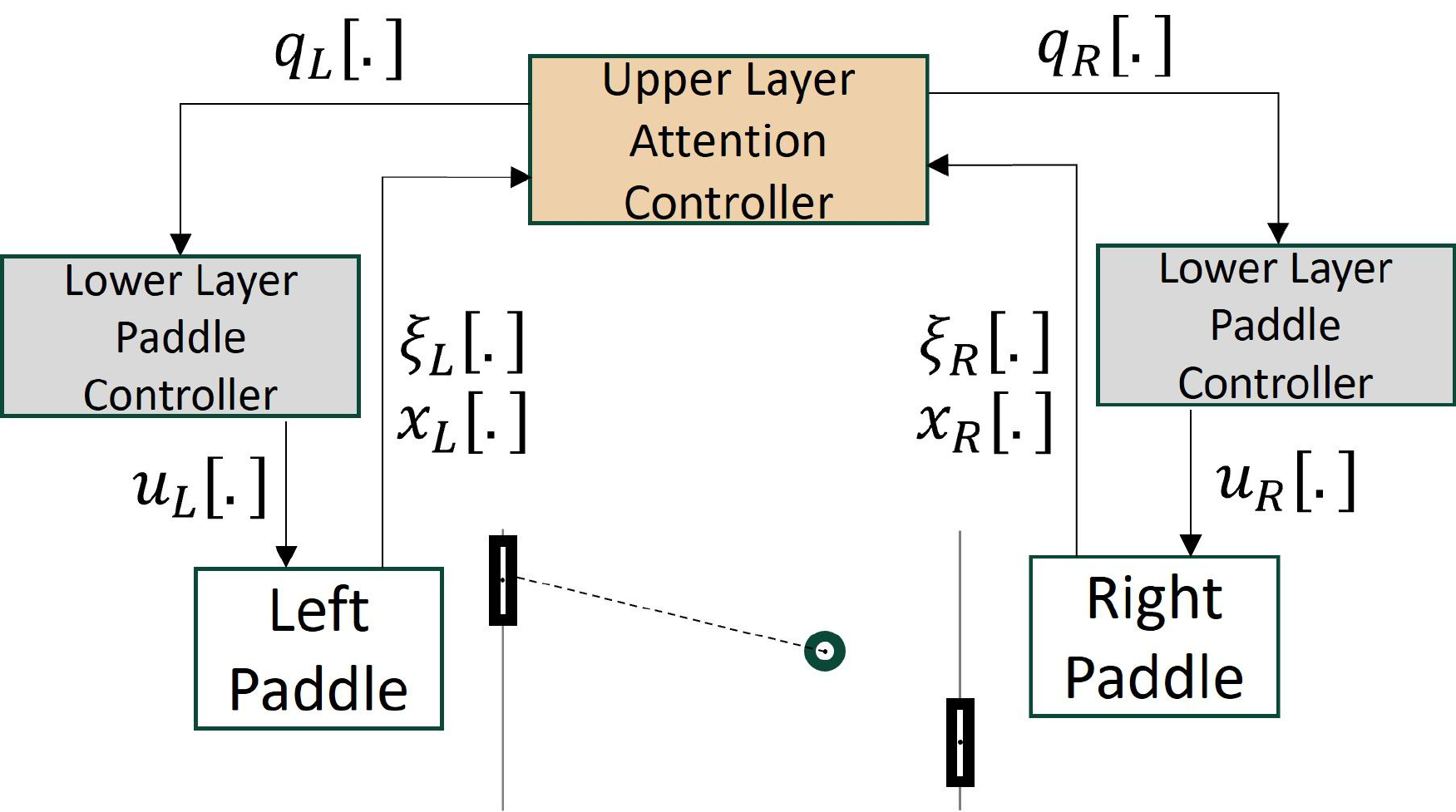}
\caption{\Red{Block diagram of two-layer optimization. Variables from Table \ref{tab:variables}.}}
\end{figure}


\Red{We present a two-layered optimization framework which provides an explicit quantification of attention that can interact with actuators, formalizing bimanual attention allocation. An upper layer solution produces an attention ``trajectory", and a lower layer produces an actuator trajectory. This model contributes broader insights into sensorimotor coordination's tradeoffs, which appear as competing objectives between tracking, coordination, and stability in the optimization problem's cost function. Convex interior-point methods are applicable; this work includes derivations of the gradients used for finding optimal solutions.} We apply this layered optimization in a single-player ``Pong" scenario where an agent tries to stabilize the ball while minimizing attention. \Red{In simulation,} we evaluate the two-layer controller's ability to converge to a steady state under various parameters, \Red{allowing} us to observe emergent behavior from trading off between attention, task difficulty, control effort, and actuator asymmetry. \Red{Factors that increase attention cost are high asymmetry, weak coordination, and aggressive stabilizing. The cost function most stable and robust to initial conditions is one that penalizes distance from the center.}

\section{Pong Model}
\begin{table}
\centering
\caption{Variables}
\begin{tabular}{|c|c|}

    \hline
    \multicolumn{2}{| c |}{Ball Variables}\\
    \hline
    $N \in \mathbb{N}_{+}$ & trajectory length (collision occurs at $t=N$)\\
    $i \in  \mathbb{N}_{+}$ & collision number/index (collisions occur at $t=N_i$)\\
    $\xi \in \mathbb{R}^4$ & ball state\\
    $\xi_v, \xi_p \in \mathbb{R}^2$ & velocity and position components of $\xi$ \\
    \hline
    \multicolumn{2}{| c |}{Paddle Variables}\\
    \hline
    $x \in \mathbb{R}^2$ & paddle state \\
    $x_{v}, x_{p}  \in \mathbb{R}$ & velocity and position component of $x$ \\
     $ \textbf{x}\in \mathbb{R}^{2(N-1)}$ & state trajectories (omit $x[1]$), $ \textbf{x} = [x[2]^\top...x[N]^\top]^\top$ \\
    $u\in \mathbb{R}$ & left/right paddle control input \\
   
    $ \textbf{u} \in \mathbb{R}^{N-1}$ & control trajectories, $\textbf{u} = [u[1]...u[N-1]]^\top$  \\
    \hline
\end{tabular}
\label{tab:variables}
\end{table}
\begin{figure}
    \centering
    \includegraphics[trim = 0cm 0.5cm 0cm 0cm, width=\linewidth]{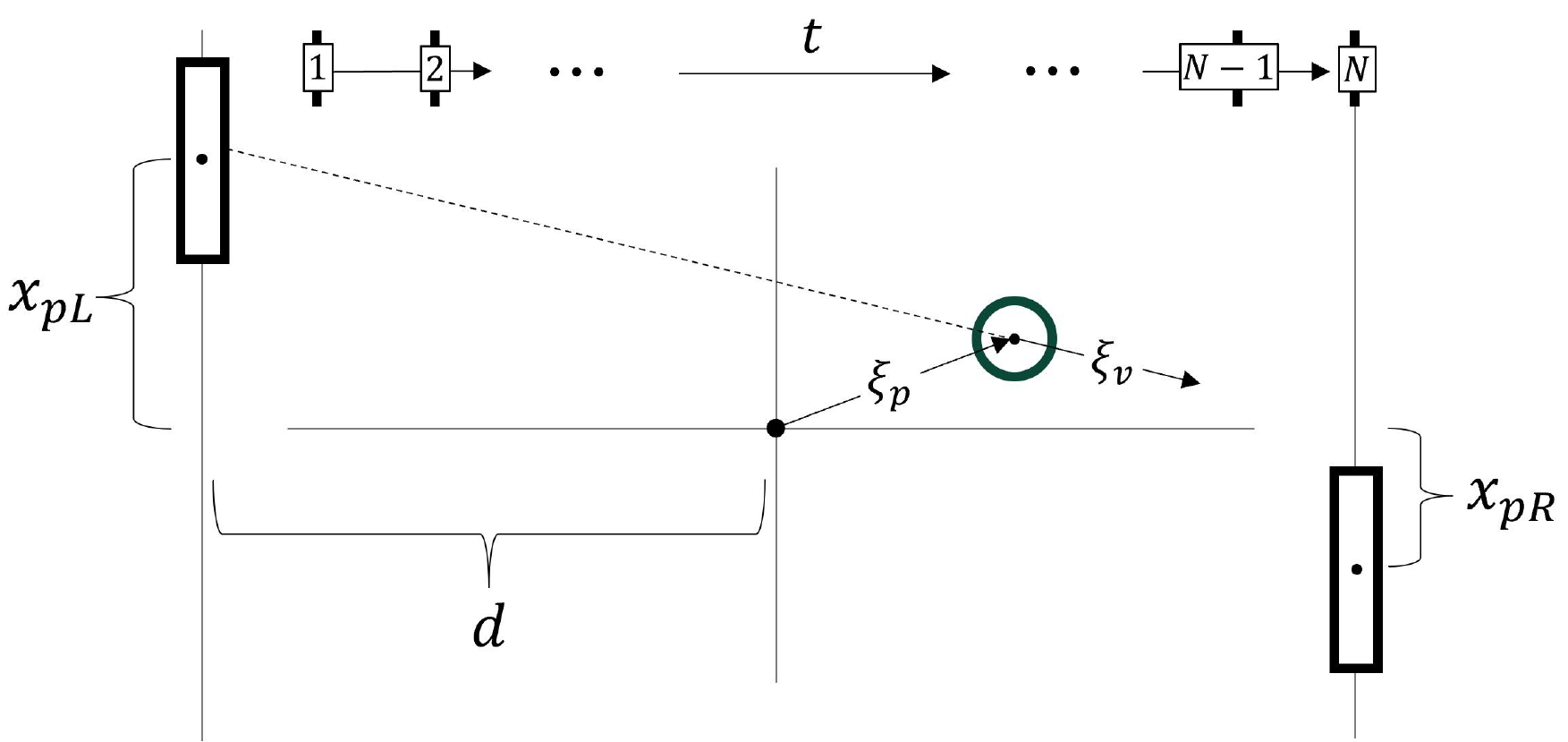}
    \caption{Diagram of the pong problem with some of the states labelled. Superscripts $p$ and $v$ indicate position and velocity respectively. Subscripts $L$ and $R$ indicate left and right respectively.}
    \label{fig:pong_vector_diagram}
\end{figure}

We explore the the relationship between attention, control effort, and bimanual tasks through a discrete-time game of pong, where each paddle is controlled by one actuator. The ball moves in a flat 2-dimensional space, while the paddles are restricted to 1-dimensional rails on both sides of the origin (See Fig.~\ref{fig:pong_vector_diagram}). This model can generalize to several setups; one could imagine a person playing an arcade machine where the actuators are the fingers pushing joysticks. A person could also be grasping hand levers, as implemented in the bimanual attention experiments in \cite{sherwoodDividedAttentionBimanual2001}, so the arms are the actuators for the paddles in a virtual environment. Alternatively, one could imagine a person juggling one object between their hands, so the actuators are the arms controlling the hand position. The hands could also be holding real paddles that hit the ball, so the actuators are the wrists. Our pong model explores strategies for dividing control effort and attention between the two actuators in all these setups. 

\subsection{Paddle Dynamics}
The paddle dynamics are modeled with a discrete-time linear system.
\begin{equation}
    x[t+1] = Ax[t] + Bu[t]
\end{equation} 
The system response is split into the zero dynamics $\textbf{A} \in \mathbb{R}^{2(N-1) \times 2}$ and input response $\textbf{B} \in \mathbb{R}^{2(N-1)\times (N-1)}$:
\begin{equation}\label{eqn:system_rollout}
\textbf{x} = \textbf{A}x[1] + \textbf{Bu}
\end{equation}
\begin{equation}
\textbf{A} = \begin{bmatrix} A \\ A^2 \\ \vdots \\ A^{N-1} \end{bmatrix}
\end{equation}
\begin{equation}
\textbf{B} = \begin{bmatrix} B \\ AB & B \\ A^2B & AB & B \\ \vdots & \ddots& \ddots& \ddots\\ A^{N-2}B & ... & ... & ... & B \end{bmatrix}
\end{equation}

\subsection{Ball Dynamics}

We represent the state of the ball with $\xi = [\xi_v^\top \hspace{0.2cm} \xi_p^\top]^\top$, where $\top$ denotes the transpose and $\xi_v, \xi_p \in  \mathbb{R}^2$ are the velocity and position components of the ball respectively. The ball collides with the paddle at time steps $t = N_i$ where $i \in \mathbb{N}^+$ indexes the collisions. 

The ball's horizontal speed is constant, so the number of time steps between collisions ($N \triangleq N_{i+1}-N_i$) is constant. The ball's vertical velocity is constant except at the collision point, where it is affected by the paddle velocity. This behavior results in (\ref{eqn:collision_to_collision}) and (\ref{eqn:on_collision}), where  (\ref{eqn:collision_to_collision}) describes the ball's movement between collisions (transition from $N_{i}$ to $N_{i+1}$), and (\ref{eqn:on_collision}) describes the ball's movement on collision (transition from $N_i$ to $N_i + 1$). For readability, equation (\ref{eqn:on_collision}) is rewritten as equation (\ref{eqn:on_collision_split}), separating velocity and position components. $I$ is an identity matrix of compatible size. 
\begin{equation}
\label{eqn:collision_to_collision}
\xi[N_{i+1}] = \begin{bmatrix} I & 0 \\ NI & I \end{bmatrix}\xi[N_i+1]
\end{equation}
\begin{equation}
\label{eqn:on_collision}
\xi[N_i+1] = \begin{bmatrix} 1 & 0 & 0 & 0\\ 0 & -1 & 0 & 0\\1 & 0 & 1 & 0\\ 0 & -1&0&1 \end{bmatrix}\xi[N_i] + \begin{bmatrix} 1 \\ 0 \\ 1 \\ 0 \end{bmatrix}x_{v}[N_i]
\end{equation}


\begin{equation}\label{eqn:on_collision_split}
\begin{array}{c}
\xi_v[N_i+1] = \begin{bmatrix} 1 & 0 \\ 0 & -1 \end{bmatrix}\xi_v[N_i] + \begin{bmatrix} x_{v}[N_i] \\ 0 \end{bmatrix} \\[0.5cm]
\xi_p[N_i+1] = \xi_p[N_i] + \xi_v[N_i+1]
\end{array}
\end{equation}

For the rest of this work, all trajectories start at $t=1$, which can generalize to any $t=N_i+1$. 


\section{Lower Controller: Paddle Tracking}
We use two lower controllers, corresponding to the left and right paddles. The only goal of the lower controller is to hit the ball with the paddle. This goal can be represented as a trajectory tracking linear-quadratic program (\ref{eqn:verbose_optimization}), where the paddle is tracking the ball's vertical position. $x_d$ is the desired state variable, and $\overline{x} \triangleq x-x_d$ is tracking error. The time-varying matrices $Q[t]$, including a terminal $Q[N]$, penalize the error $\overline{x}$. The control penalty $r$ is constant. We denote $\mathcal{D}^n$ as the set of all $n \times n$ positive semidefinite diagonal matrices. The vector $x[1]$ is initialized to a constant $x_o$.
\begin{equation}\label{eqn:verbose_optimization}
\begin{array}{rrcl}
\displaystyle \min_{u[1]...u[N-1]} & \multicolumn{3}{l}{\mathcal{J}_P = \overline{x}[N]^\top Q[N]\overline{x}[N]} \\ 
& \multicolumn{3}{l}{\displaystyle + \sum_{\uptau=1}^{N-1} \overline{x}[\uptau]^\top Q[\uptau]\overline{x}[\uptau] + {u}[\uptau]^2r}
\\[0.5cm]
\textrm{s.t.} & \multicolumn{3}{l}{x[t+1] = Ax[t] + Bu[t]} \\[0.1cm]
& \multicolumn{3}{l}{Q[t] \in \mathcal{D}^2} \\[0.1cm]
& \multicolumn{3}{l}{r > 0} \\[0.1cm]
& \multicolumn{3}{l}{x[1] = x_o} \\[0.1cm]
\end{array}
\end{equation}
The problem formulation (\ref{eqn:verbose_optimization}) is equivalent to the more compact representation (\ref{eqn:compact_optimization}), using the variables from (\ref{eqn:system_rollout}). \textbf{x}$_d$ is the desired trajectory and \textbf{Q} is a diagonal matrix with the matrices $Q[2] \hspace{0.1cm} ... \hspace{0.1cm} Q[N]$ on its diagonal. The cross-term \textbf{$-2$x$_d^\top$Qx} comes from distributing the quadratic error cost (the constant \textbf{x$_d^\top$Qx$_d$} does not impact the solution).
\begin{equation} \label{eqn:compact_optimization}
\begin{array}{rrclcl}
\displaystyle \min_{\textbf{u}} & \multicolumn{3}{l}{\mathcal{J}_P = \textbf{x$^\top$Qx} - 2\textbf{x$_d^\top$Qx} +  r\textbf{u}^\top\textbf{u}} \\
\textrm{s.t.} & \textbf{x} = \textbf{A}x[1] + \textbf{Bu} \\[0.1cm]
& \multicolumn{3}{l}{\textbf{Q} \in \mathcal{D}^{2(N-1)}} \\[0.1cm]
& \multicolumn{3}{l}{r > 0} \\[0.1cm]
& \multicolumn{3}{l}{x[1] = x_o} \\[0.1cm]
\end{array}
\end{equation} 
The explicit solution to this program will be used to solve the upper controller. Taking the derivative of the cost $\mathcal{J}_P$, setting it to zero, and solving for $\textbf{u}$ results in the following:
\begin{equation}\label{eqn:closed_form_u}
\begin{array}{c}
\textbf{u} = (rI + \textbf{B$^\top$QB})^{-1}\textbf{B$^\top$Zq}\\[0.1cm]

\textbf{z} \triangleq \textbf{x}_d - \textbf{A}x[1] \triangleq diag(\textbf{Z})\\[0.1cm]

\textbf{q} \triangleq [q_v[2], \hspace{0.1cm} q_p[2],\hspace{0.1cm} \hdots q_v[N],\hspace{0.1cm} q_p[N]]^\top\triangleq diag(\textbf{Q})
\end{array}
\end{equation}

\textbf{Z} and \textbf{Q} are diagonal matrices, where the values on the diagonals are the vectors \textbf{z} and \textbf{q} respectively. The values in \textbf{z} and \textbf{Z} are the tracking errors when \textbf{u} is the zero vector. 
\section{Upper Controller: Attention Allocation}
The upper controller goal is to minimize attention while rallying the ball. Consider the case of the ball headed towards a stationary paddle. The ball will collide with the paddle without any control effort or attention. If we increase the attention used in this task, the ball will still be hit with no control effort. Since the task can succeed with or without attention, any energy spent on processing visual information and computing an optimal trajectory is wasted. Now consider the case where the paddle is not blocking the ball's trajectory. If the agent is not paying any attention (e.g. their eyes are closed, they get distracted, they are too tired), the paddle will miss the ball and the task fails. Therefore, some amount of attention must be spent in order for the task to succeed. In this work, we quantify attention with the vector \textbf{q} as defined in (\ref{eqn:closed_form_u}). Setting \textbf{q} as the zero vector produces the no-attention behavior, while large values of \textbf{q} allow the task to succeed.

\begin{figure}
\includegraphics[trim = 0cm 0.5cm 0cm 0cm,width=.95\linewidth]{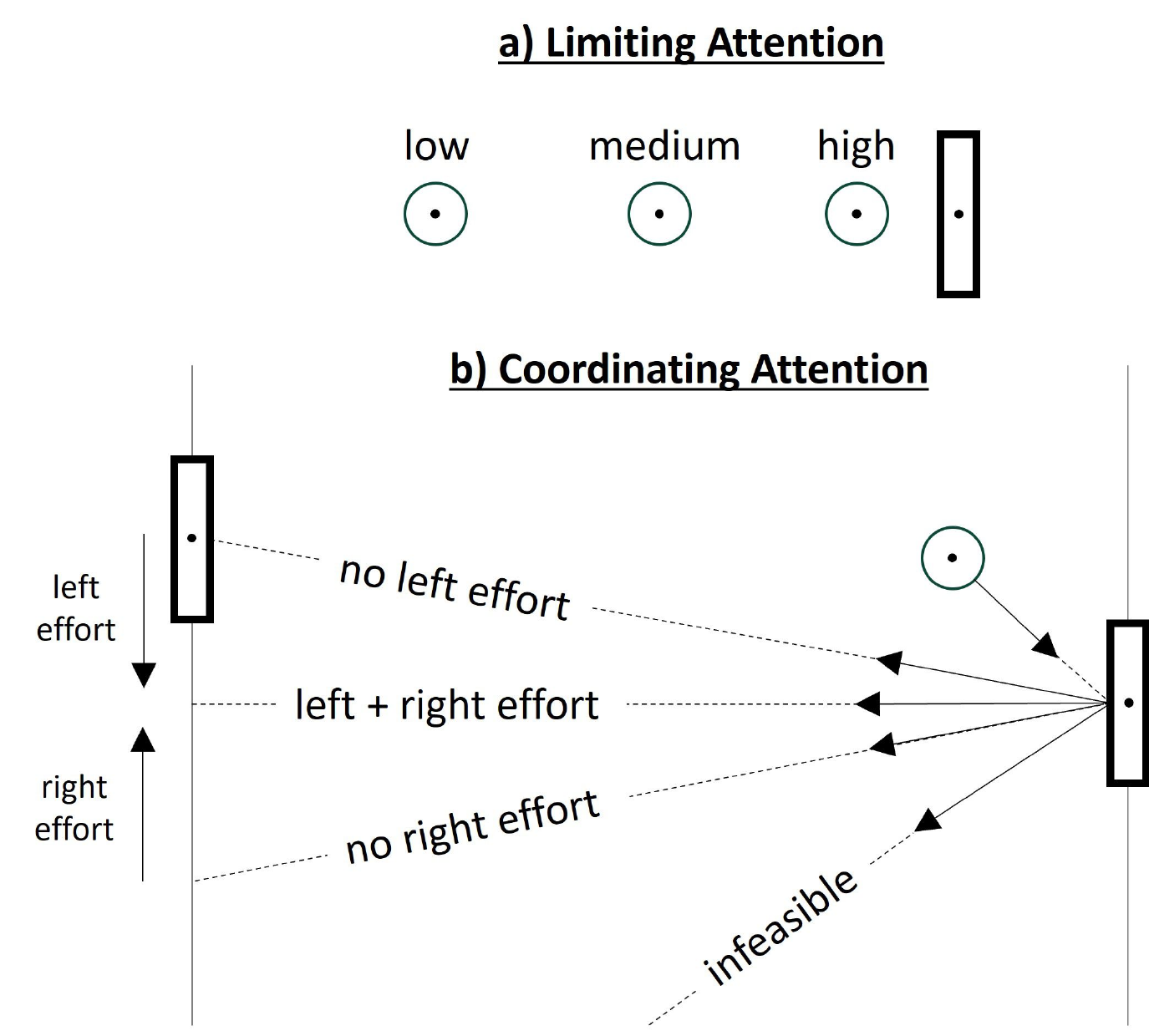}
\caption{Two strategies for reducing attention. Attention can be limited when the ball is far away from the paddle. Attention can also be saved by manipulating the ball's reflection angle towards the opposite paddle.}
\label{fig:att_strat}
\end{figure}

 There are two ways to reduce the overall attention. First, do not track the ball when it is far away from the paddle. Second, hit the ball towards the opposite paddle so that the ball is easier to reach on the next collision (Fig. \ref{fig:att_strat}). These two strategies can be represented as two separate optimization problems, which we will call the tracking attention problem and coordination attention problem. These two problems are combined for the full bimanual task. The following sections will derive the costs and constraints for both problems.

\subsection{Tracking Attention Problem}

When a paddle is trying to reach the ball in time, it is only concerned with the position components of the desired trajectory $\textbf{x}_d \in \mathbb{R}^{2(N-1)}$, which we will call $\textbf{x}_{p,d} \in \mathbb{R}^{N-1}$. Since the goal of the lower controller is to track the ball's position, $\textbf{x}_{p,d}$ takes its values from the vertical positions of $\xi_p[t]$. $e_i$ is the $i$-th standard basis vector (therefore $e_1^\top\xi_p$ is vertical position and $e_2^\top\xi_p$ is horizontal position).
\begin{equation}
\textbf{x}_{p,d} = \begin{bmatrix} e_1^\top\xi_p[2], & e_1^\top\xi_p[3], & \hdots & e_1^\top\xi_p[N] \end{bmatrix}^\top
\end{equation}
Similarly, $\textbf{q}_{p} \in \mathbb{R}_{\geq0}^{N-1}$ corresponds to the position error penalties in \textbf{q}. The tracking attention problem is written as:
\begin{equation}\label{eqn:position_problem}
\begin{array}{rrcl}
\displaystyle \min_{\textbf{q}_p} & \multicolumn{3}{l}{\mathcal{J}_{pA} = \textbf{q}_p^\top \textbf{q}_p} \\[0.1cm]
\textrm{s.t.} & \multicolumn{3}{l}{\vert E(\textbf{q}_p) \vert \leq \upepsilon} \\[0.1cm]
& \multicolumn{3}{l}{\textbf{q}_p \in \mathbb{R}_{\geq0}^{N-1}}\\[0.1cm]
\end{array}
\end{equation}
where $E \triangleq e_1^\top\xi_p[N] - x_p[N]$ is the error when the ball is supposed to be hit ($x_p[N]$ is the final position component of \textbf{x}), and $\upepsilon > 0$ is the acceptable error margin. The value of $\upepsilon$ corresponds to the length of the paddle, since a larger paddle allows for a larger margin of error. Using (\ref{eqn:system_rollout}) and (\ref{eqn:closed_form_u}), $E$ can be expressed in terms of $\textbf{q}_p$. 
\begin{flalign*}
\hspace{0.5cm} E(\textbf{q}_p) = e_1^\top\xi_p[N] - x_p[N]&&
\end{flalign*}
\begin{equation} \label{eqn:E_to_q}
  \begin{array}{rl}
  &= e_1^\top\xi_p[N] - e_1^\top A^{N-1}x[1] \\[0.1cm]
  & \hspace{0.3cm}  - b_p[N](rI + \textbf{B$_p^\top$Q$_p$B$_p$})^{-1}\textbf{B$^\top_p$Z$_p$q$_p$} \\ [0.1cm]
  &= z_p[N] - b_p[N]^\top (rI + \textbf{B$_p^\top$Q$_p$B$_p$})^{-1}\textbf{B$^\top_p$Z$_p$q$_p$}
  \end{array}
\end{equation}

where $b_p[N]$ is the transposed row of \textbf{B}$_p$ (so $b_p[N]^\top$ is a row vector) that corresponds to the position of $x[N]$, and $z_p[N]$ is the scalar  corresponding to the position component of \textbf{z} at timestep $N$ (so $z_p[N] = e_1^\top(\xi_p[N] - A^{N-1}x[1])$). The matrices $\textbf{B}_p$, $\textbf{Q}_p$, and $\textbf{Z}_p$ are the matrices in (\ref{eqn:closed_form_u}), but with only the rows and columns relevant to position. In other words, since \textbf{q} is size $2(N-1) \times 1$ and $\textbf{q}_p$ is size $(N-1) \times 1$, the matrices in the expression (\ref{eqn:closed_form_u}) must be shrunk accordingly.

The constraint $E \leq \upepsilon$ combined with $\textbf{q}_p \in \mathbb{R}_{\geq0}^{N-1}$ constrains $\textbf{q}_p$ to a convex hyperbola in the positive orthant (Fig. \ref{fig:grad}). The optimal value of $\textbf{q}_p$ can be found using interior-point methods, which requires the gradient of $E$. Using the Woodbury Matrix Identity, the matrix inverse in (\ref{eqn:E_to_q}) can be rewritten:
\begin{equation}\label{eqn:woodbury}
\begin{array}{c}
(rI + \textbf{B$_p^\top$Q$_p$B$_p$})^{-1} =  r^{-1}I-r^{-2}\textbf{B$_p^\top$C$^{-1}$B$_p$}\\[0.1cm]
\textbf{C} = \textbf{Q$_p^{-1}$}  + r^{-1}\textbf{P}\\[0.1cm]
\textbf{P} = \textbf{B$_p$B$_p^\top$}\\
\end{array}
\end{equation}
Using this expression, the gradient can be derived.  First, distribute the elements of $E(\textbf{q}_p)$ using (\ref{eqn:woodbury}) and (\ref{eqn:E_to_q}).
\begin{flalign*}
E(\textbf{q}_p)&= z_p[N] - b_p[N]^\top (r^{-1}I-r^{-2}\textbf{B$_p^\top$C$^{-1}$B$_p$})\textbf{B$^\top_p$Z$_p$q$_p$}&&
\end{flalign*}
\begin{equation}
\begin{array}{rl}
&\hspace{-1.56cm}=z_p[N] -r^{-1}b_p[N]^\top\textbf{B$^\top_p$Z$_p$q$_p$}\\[0.1cm]
&\hspace{-1.26cm}+r^{-2}b_p[N]^\top\textbf{B$_p^\top$C$^{-1}$P}\textbf{Z$_p$q$_p$}\\[0.1cm]
\vspace{0.15cm}
&\hspace{-1.56cm}= z_p[N] - r^{-1}\textbf{k}_0^\top\textbf{k}_1+ r^{-2}\textbf{k}_0^\top\textbf{K}\textbf{k}_1\\[0.1cm]
\textbf{k}_0&\triangleq\textbf{B$_p$}b_p[N] \in \mathbb{R}^{(N-1)\times1}\\[0.1cm]
\textbf{k}_1&\triangleq\textbf{Z$_p$q$_p$} \hspace{.47cm} \in \mathbb{R}^{(N-1)\times1}\\[0.1cm]
\textbf{K}&\triangleq\textbf{C$^{-1}$P}\hspace{.39cm} \in \mathbb{R}^{(N-1)\times(N-1)}\\[0.1cm]
\end{array}
\end{equation}
We write derivatives with the following convention. If the derivative argument is a vector, the result is a Jacobian matrix ($\frac{d}{d\textbf{q}_p}\textbf{k}_1=\textbf{Z}_p$). If the argument is a scalar, the result is a column vector ($\frac{d}{d\textbf{q}_p}(\textbf{k}_0^\top\textbf{k}_1) = \textbf{Z}_p\textbf{k}_0$). \textbf{K} and \textbf{k}$_1$ depend on \textbf{q}$_p$, but \textbf{k}$_0$ does not. Now we must find the dervative of the $\textbf{k}_0^\top\textbf{K}\textbf{k}_1$ term from above.  Using product rule and the matrix inverse derivative property:
\begin{flalign*}
\frac{d}{d\textbf{q}_p}(\textbf{k}_0^\top\textbf{K}\textbf{k}_1) &= \frac{d}{d\textbf{q}_p}(\textbf{k}_1)^\top\textbf{K}^\top\textbf{k}_0 
+ \textbf{k}_1^\top\frac{d}{d\textbf{q}_p}(\textbf{K}^\top\textbf{k}_0)&&
\end{flalign*}
\begin{equation}\label{eqn:matrix_derivative}
\begin{array}{rl}
&\hspace{6pt}= \textbf{Z$_p$}\textbf{K}^\top\textbf{k}_0 + \textbf{k}_1^\top\textbf{P}\frac{d}{d\textbf{q}_p}(\textbf{C}^{-1}\textbf{k}_0)\\[0.1cm]
&\hspace{6pt}= \textbf{Z$_p$}\textbf{K}^\top\textbf{k}_0 - \textbf{k}_1^\top\textbf{PC}^{-1}\frac{d}{d\textbf{q}_p}(\textbf{C})\textbf{C}^{-1}\textbf{k}_0
\end{array}
\end{equation}
\Red{In (\ref{eqn:matrix_derivative}) there is a derivative with a matrix argument ($\frac{d}{d\textbf{q}_p}\textbf{C} = \frac{d}{d\textbf{q}_p}\textbf{Q}_p^{-1}$). Since \textbf{Q}$_p$ is diagonal with only one unique variable per diagonal component, this derivative works as an element-wise operation along the diagonal (the elements of \textbf{q}$_p$). Therefore, $\frac{d}{d\textbf{q}_p}\textbf{Q}_p^{-1}$ is a matrix with the diagonal -\textbf{q}$_p^{-2} \triangleq [-q_p[2]^{-2} \hdots  -q_p[N]^{-2}]^\top$. This multivariable derivative is abusing the univariable matrix inverse derivative notation.}

Combining all the gradients gives us the following, where $\odot$ is element-wise vector multiplication:
\begin{flalign*}
\hspace{0.7cm}\displaystyle \frac{dE}{d\textbf{q$_p$}} &\displaystyle=-r^{-1}\frac{d}{d\textbf{q}_p}(\textbf{k}_0^\top\textbf{k}_1) +r^{-2}\frac{d}{d\textbf{q}_p}(\textbf{k}_0^\top\textbf{K}\textbf{k}_1)&&
\end{flalign*}
\begin{equation}
\begin{array}{rl}
             &= -r^{-1}\textbf{Z}_p\textbf{k}_0 + r^{-2}\textbf{Z$_p$}\textbf{K}^\top\textbf{k}_0\\[0.1cm]
             &\hspace{0.3cm}\displaystyle-r^{-2}\textbf{k}_1^\top\textbf{PC}^{-1}\frac{d}{d\textbf{q}_p}(\textbf{Q}_p^{-1})\textbf{C}^{-1}\textbf{k}_0\\[0.1cm]
             & = -r^{-1}\textbf{Z$_p$B$_p$}b_p[N] \\[0.1cm]
             & \hspace{0.3cm} + r^{-2}\textbf{Z$_p$PC}^{-1}\textbf{B$_p$}b_p[N] \\[0.1cm]
             & \hspace{0.3cm} -r^{-2}(-\textbf{q$_p^{-2} \odot$ 
                                                                  C}^{-1}\textbf{P Z$_p$q$_p$ $\odot$ C}^{-1}\textbf{B$_p$}b_p[N]) \\
\end{array}
\end{equation}
\subsection{Coordination Attention Problem}
Since the coordination problem requires both paddles, the subscripts $L$ and $R$ will be used to distinguish variables corresponding to the left and right sides. In the following sections, the ball will start on the left side and travel towards the right paddle. The ball hits the right paddle at timestep $N_i$ and the left paddle at timestep $N_{i+1}$.

The reflection angle off the right paddle can be influenced by the velocity of the paddle at the time of collision, as described by (\ref{eqn:on_collision_split}). \Red{If the ball is reflected towards the left paddle, the left attention is minimized, but at the cost of right attention (Fig.~\ref{fig:att_strat}), establishing a tradeoff in the cost function.} We assume the left paddle does not move between the collision points, so $x_{pL}[N] = x_{pL}[2N]$. Let $v_d$ be the velocity of the right paddle at the collision point $t=N$ such that the ball reaches the left paddle's position at $t=2N$, so $x_{pL}[2N] = e_1^\top\xi_{p}[2N]$. This velocity $v_d$ can be used as a tracking point in the terminal cost of the right lower controller (\ref{eqn:verbose_optimization}).
\begin{flalign*}
\overline{x}_R[N]^\top Q_R[N]\overline{x}_R[N] & = q_{vR}[N](x_{vR}[N]-v_d)^2&&
\end{flalign*}
\begin{equation}
\begin{array}{rl}
& \hspace{2cm} + q_{pR}[N](x_{pR}[N]-e_1^\top\xi_p[N])^2
\end{array}
\end{equation}
Increasing the penalty $q_{vR}[N]$ will decrease the error $x_{vR}[N]-v_d$, which then decreases the attention \textbf{q$_{pL}^\top$q$_{pL}$} (starts at $N+1$, \textbf{q$_{pL}$} $\triangleq [q_{pL}[N+1] \hdots q_{pL}[2N]]^\top$). When $q_{vR}[N]$ increases, the feasible region of the left lower controller (\ref{eqn:position_problem}) grows. Similarly, decreasing $q_{vR}[N]$ will increase \textbf{q$_{pL}^\top$q$_{pL}$} and shrink the left feasible region. The coordination problem optimizes the tradeoff between the left and right attention. 
\begin{equation}\label{eqn:velocity_problem_1}
\begin{array}{rrcl}
\displaystyle \min_{q_{vR}[N]} & \multicolumn{3}{l}{\mathcal{J}_{vA} = q_{vR}[N]^2 + \textbf{q$_{pL}^\top$q$_{pL}$}} \\[0.1cm]
\textrm{s.t.} & \multicolumn{3}{l}{q_{vR}[N] \geq 0}  \\
\end{array}
\end{equation}
This problem statement suggests that \textbf{q$_{pL}$} has a closed-form expression in terms of $q_{vR}[N]$ from the solution of (\ref{eqn:position_problem}). However, we can use an approximation \textbf{q$_{pL}^\top$q$_{pL}$} $\approx \hat{q}_{pL}^2 \triangleq {q}_{pL}[2N]^2$ instead. This is a reasonable approximation, since the solutions of (\ref{eqn:position_problem}) result in $q_p[N]$ being significantly larger than $q_p[N-1], ...,q_p[2]$ (see Section \ref{sec:results} for more details). An expression for $\hat{q}_{pL}$ can be found in terms of $q_{vR}[N]$ by setting the cost of (\ref{eqn:position_problem}) to $\mathcal{J}_{pA} = q_{pL}[2N]^2$, so the problem can be solved along one dimension.

We first introduce the following variables:
\begin{equation}
\begin{array}{c}
v_o = e_2^\top A^{N-1}x_R[1]\\[0.1cm]
\displaystyle \gamma_L = \frac{r_L}{k_{pL}} \hspace{0.5cm} \gamma_R = \frac{r_R}{k_{vR}} \\[0.4cm]
k_{pL} = b_p[2N]^\top b_p[2N] \hspace{0.5cm} k_{vR} = b_v[N]^\top b_v[N]\\
\end{array}
\end{equation}
$v_o$ is the collision velocity of the right paddle if \textbf{u} is the zero vector. $\gamma_L$ and $\gamma_R$ are constants that appear when the expressions (\ref{eqn:closed_form_u})(\ref{eqn:E_to_q}) are compressed to one dimension. We use (\ref{eqn:E_to_q}) to start with an initial expression for $\hat{q}_{pL}$, then substitute until the expression is in terms of $q_{vR}[N]$. We set the constraint $E \leq \upepsilon$ as an equality (assume $E > 0$ without loss of generality). This gives us a way to solve for $\hat{q}_{pL}$:
\begin{equation*}
\begin{array}{rl}
\hspace{-2.1cm}E(\hat{q}_{pL}) &= z_{pL}[2N](1+\gamma_L^{-1}\hat{q}_{pL})^{-1} = \upepsilon\\[0.1cm]
\hspace{-2.1cm}\hat{q}_{pL} &\displaystyle= \gamma_L(\upepsilon^{-1}z_{pL}[2N]-1)\\[0.3cm]
\hspace{-2.1cm}z_{pL}[2N]& \triangleq e_1^\top\xi_p[2N] - x_{pL}[N]\\[0.1cm]
\end{array}
\end{equation*}
\begin{equation}\label{eqn:qL_first_step}
\hspace{1cm}= e_1^\top\xi_p[N] + N(e_1^\top\xi_v[N] + x_{vR}[N]) - x_{pL}[N]
\vspace{0.3cm}
\end{equation}
$z_{pL}[2N]$ is simply the scalar version of \textbf{z}. By setting $z_{pL}[2N] = 0$ and $x_{vR}[N]=v_d$, we can solve for $v_d$.
\begin{equation}\label{eqn:v_des}
v_d = \frac{1}{N}(x_{pL}[N]-e_1^\top\xi_p[N]) - \xi_v[N] 
\end{equation}
We must find $z_{pL}[2N]$ in terms of $q_{vR}[N]$ through $x_{vR}[N]$. By doing the same substitution in (\ref{eqn:E_to_q}) for one dimension, we get an explicit expression for $x_{vR}[N]$.
\begin{equation}\label{eqn:qL_last_step}
\begin{array}{rl}
x_{vR}[N] &= \mu v_d + (1-\mu)v_o \\[0.1cm]
\mu &\displaystyle =\frac{q_{vR}[N]}{\gamma_R+q_{vR}[N]}\\
\end{array}
\end{equation}
Note that $\mu \in [0, 1)$. Combining (\ref{eqn:qL_first_step})-(\ref{eqn:qL_last_step}) gives us $\hat{q}_{pL}$ in terms of $q_{vR}[N]$ (through $\mu$).
\begin{equation}\label{eqn:qL_estimate}
\begin{array}{rl}
\hat{q}_{pL} &= \displaystyle\gamma_L(\upepsilon^{-1}(1-\mu)(\kappa_1+N\kappa_2)-1)\\[0.1cm]

\kappa_1 &=e_1^\top\xi_p[N]-x_{pL}[N]\\[0.1cm]
\kappa_2 &=e_1^\top\xi_v[N]+v_o\\
\end{array}
\end{equation}
Rewriting the coordination problem (\ref{eqn:velocity_problem_1}) by incorporating the expression (\ref{eqn:qL_estimate}) results in the following:
\begin{equation}\label{eqn:velocity_problem_2}
\begin{array}{rrcl}
\displaystyle \min_{q_{vR}[N]} & \multicolumn{3}{l}{\mathcal{J}_{vA} = q_{vR}[N]^2 + \hat{q}_{pL}^2} \\[0.1cm]
\textrm{s.t.} & \multicolumn{3}{l}{\hat{q}_{pL} = \displaystyle\gamma_L(\upepsilon^{-1}(1-\mu)(\kappa_1+N\kappa_2)-1)}  \\[0.1cm]
& \multicolumn{3}{l}{q_{vR}[N], \hat{q}_{pL} \geq 0}  \\
\end{array}
\end{equation}
Combining the tracking problem (\ref{eqn:position_problem}) and coordination problem (\ref{eqn:velocity_problem_2}) yields the overall attention optimization problem. \textbf{q}$_R$ is a vector representing attention for the right paddle, while $\hat{q}_{pL}$ is the attention for the left (switch the $R, L$ subscripts when the ball travels left). \textbf{q}$_R$ includes both $\textbf{q}_{pR}$ and $q_{vR}[N]$. Note that $E$ now depends on $q_{vR}[N]$ in addition to $\textbf{q}_p$, which adds a dimension to (\ref{eqn:closed_form_u}). $\mathcal{J}_{pA}$ is the cost from the tracking attention problem, and $\mathcal{J}_{vA}$ is the cost for the coordination problem. $E$ is the error at the time of collision, which is constrained below the margin $\upepsilon$. This constraint limits the solutions to a convex region (See Fig.~\ref{fig:grad}). 
\begin{equation}\label{eqn:attention_problem_full}
\begin{array}{rrcl}
\displaystyle \min_{\textbf{q}_R} & \multicolumn{3}{l}{\mathcal{J}_{A} = \mathcal{J}_{pA} + \mathcal{J}_{vA}} \\[0.1cm]
\textrm{s.t.} & \multicolumn{3}{l}{\vert E(\textbf{q}_{R}) \vert \leq \upepsilon} \\[0.1cm]
&\multicolumn{3}{l}{\hat{q}_{pL} = \displaystyle\gamma_L(\upepsilon^{-1}(1-\mu)(\kappa_1+N\kappa_2)-1) \geq 0}\\[0.1cm]

& \multicolumn{3}{l}{\textbf{q}_R \in \mathbb{R}_{\geq0}^{2(N-1)}}\\[0.1cm]
\end{array}
\end{equation}
\section{Simulations and Results}\label{sec:results}
All simulations are done in MATLAB. The lower paddle problem (\ref{eqn:compact_optimization}) is solved with YALMIP \cite{Lofberg2004}. A custom interior point method was used to find solutions to the upper attention problem (\ref{eqn:attention_problem_full}), \Red{as commercial solvers did not accept expression (\ref{eqn:E_to_q}).}  Since most of the values earlier in the time horizon will be close to 0, we solve the attention problem in four dimensions ($N$ to $N-3$), which yields similar solutions to the solutions over the entire horizon. For all examples, $B = [1 \hspace{0.2cm} 0]^\top$, the horizontal speed of the ball is 1.3m/s, and the paddles are 1.3m apart, which results in $N=10$. The linear paddle dynamics are $A = e^{A'}, A' = \begin{bmatrix} -\delta & 0\\ 1 & 0\end{bmatrix}$, where $\delta$ is a damping coefficient. The error margin $\upepsilon$ is 0.03m. If the margin was set to the exact paddle length, the controller would only \Red{hit the ball with the paddle edge}. Therefore, the margin is set to a value smaller than the length of a realistic paddle. $r$ is fixed at 1; changing $r$ simply scales the attention needed for a task to succeed.

\Red{\subsection{Simulation Considerations}
An initial feasible point of \textbf{q} is easy to find because the feasible region opens up to infinity, so arbitrarily large values of \textbf{q} will likely be feasible unless the problem, and therefore the whole space, is infeasible (e.g. ball is out of reach). \textbf{Q}$_p^{-1}$ in Equation (\ref{eqn:woodbury}) presents a numerical difficulty. The values of \textbf{q}$_p$ quickly approach 0 and make \textbf{Q}$_p$ singular. While this can be prevented by amplifying the orthant constraint, which increases the gradient magnitude near the constraint edge, the solutions are significantly worse. The gradient's magnitude exhibits an all-or-nothing behavior, almost vanishing in most places and growing several orders of magnitude near the constraint edges. Therefore, the optimization step size must be adjusted adaptively.} 

\subsection{Tracking Attention, Single Collision}
\begin{figure}
\includegraphics[trim = 0cm 0.7cm 0cm 0cm, width=0.9\linewidth]{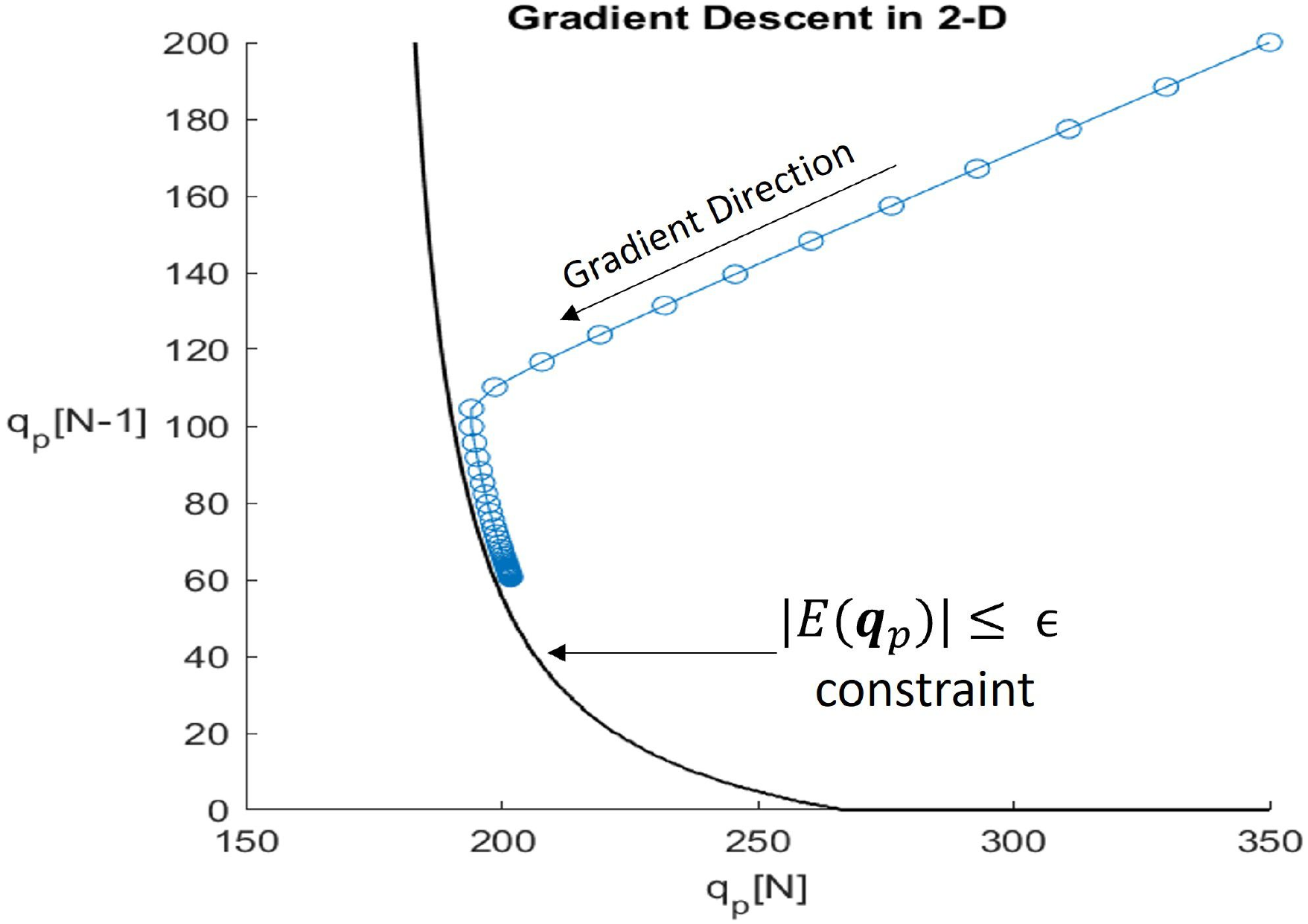}
\caption{Solving (\ref{eqn:position_problem}) with gradient descent, where \textbf{q}$_p$ is restricted to two dimensions ($N$ and $N-1$). This shows that the constraint is a convex hyperbola in the positive orthant. The value at $N-1$ is allowed to be much lower than the value at $N$.}
\label{fig:grad}
\end{figure}

First, we will study the tracking problem (\ref{eqn:position_problem}), a sub-problem of the full bimanual coordination problem (\ref{eqn:attention_problem_full}). If the goal is to minimize attention while successfully hitting the ball, we would expect attention to be allocated towards the end of the ball's trajectory. Fig.~\ref{fig:grad} displays the feasible region in two dimensions. Naturally, the constraint allows low values at $N-1$ but only high values at $N$. 

When the optimization problem is extended over the whole trajectory of the ball, the controller must decide how early in the ball's trajectory it should start adding attention. While information about the ball's state is 
irrelevant when the ball is far from the paddle, allocating all the attention to the last time step increases the chance of missing the ball, so there must be some window of time before the collision when the ball state is tracked. Taking this into account, the optimal solution lies in between the extreme cases of not paying enough attention and paying too much attention. Furthermore, we would expect this solution to pay attention earlier for ``Hard" tasks (ball is hard to reach) and later for ``Easy" tasks (ball travels towards the paddle).

\begin{figure}
\includegraphics[trim = 0cm 1cm 1cm 0cm, width=0.95\linewidth]{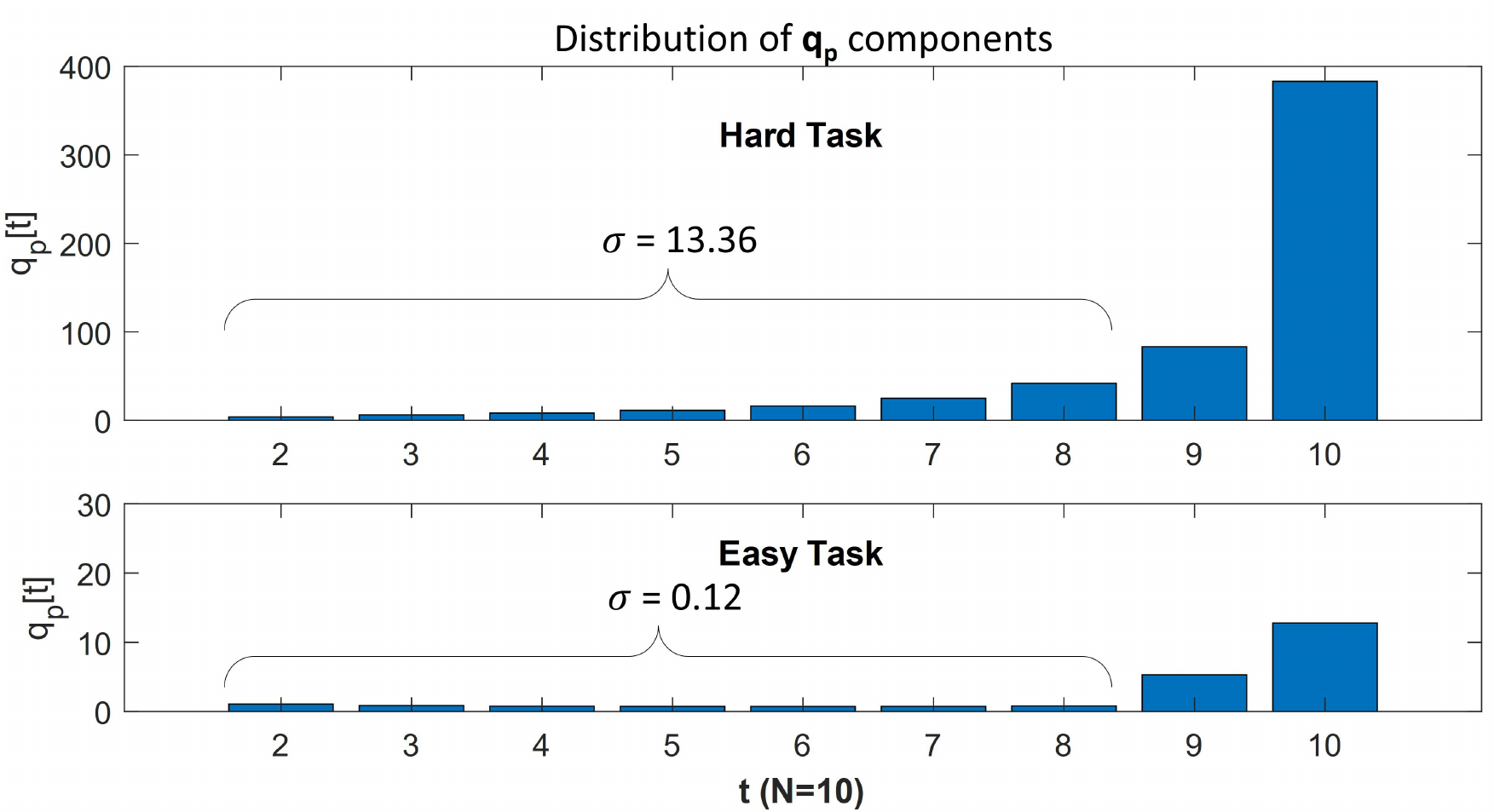}
\caption{\Red{Attention per timestep ($q_p[t]$) after solving (\ref{eqn:position_problem}). Hard tasks have a larger standard deviation $\sigma$ across time, indicating an attention window that extends further back. In the Easy task, fixing the attention values to 0 for $t=2...8$ does not change the solution. The total attention costs $\mathcal{J}_{pA}$ for the Hard and Easy task are 39748 and 192 respectively.}}
\label{fig:q_distro}
\end{figure}

Fig.~\ref{fig:q_distro} shows the time distribution of attention for an Easy and Hard task. As expected, large values of attention extend further back in time for the Hard task, in contrast to the Easy task where attention drops to a steady near-zero value after only two time steps before the collision. If we take the Hard task and constrain it to only one dimension at $q_p[N]$ (e.g. the controller only pays attention at the last second), the total cost $\mathcal{J}_{pA}$ is 72361, which is significantly greater than the cost of the \Red{multi-dimension} Hard task ($\mathcal{J}_{pA}$=39748). This supports the intuition that compressing too much attention towards the end of the ball's trajectory is sub-optimal, \Red{and the task would fail with a strict attention upper bound.}

\subsection{Coordination Attention, Multi-Collision}
\begin{figure*}
\includegraphics[trim = 0cm 0.7cm 0cm 0cm, width=1\linewidth]{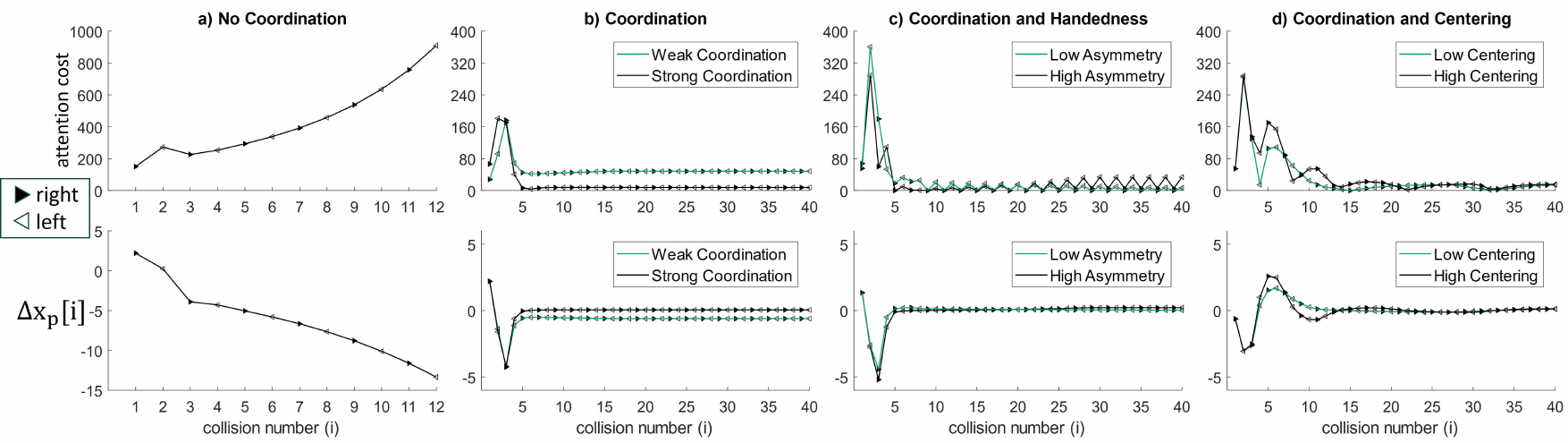}
\caption{\Red{The attention cost \textbf{q}$[i]_p^\top$\textbf{q}$[i]_p$ and paddle movement $\Delta x_p[i] \triangleq x_p[N_i] - x_p[N_{i-1}+1]$ over 40 collisions (12 for \textit{a)}). If the ball and paddle position converge, $\Delta x_p[i]$ should approach 0. Converging attention does not guarantee $\Delta x_p[i]$ approaches 0. \textit{a)} No Coordination: Attention cost and paddle movement diverge. \textit{b)} Coordination: attention converges in both strong and weak cases, but the weak case drifts in steady-state ($\Delta x_p[i]$ approaches nonzero value). \textit{c)} Handedness: An attention gap appears between the right and left actuators. Attention converges, but the high asymmetry case cannot eliminate drift as well as low asymmetry. \textit{d)} Centering: Attention converges slowly compared to \textit{b)} and \textit{c)}, but drift is eliminated. 
}}
\label{fig:att_u_plots}
\end{figure*}

\begin{figure}
\includegraphics[trim = 0cm 1cm 2cm 1cm, scale=0.28]{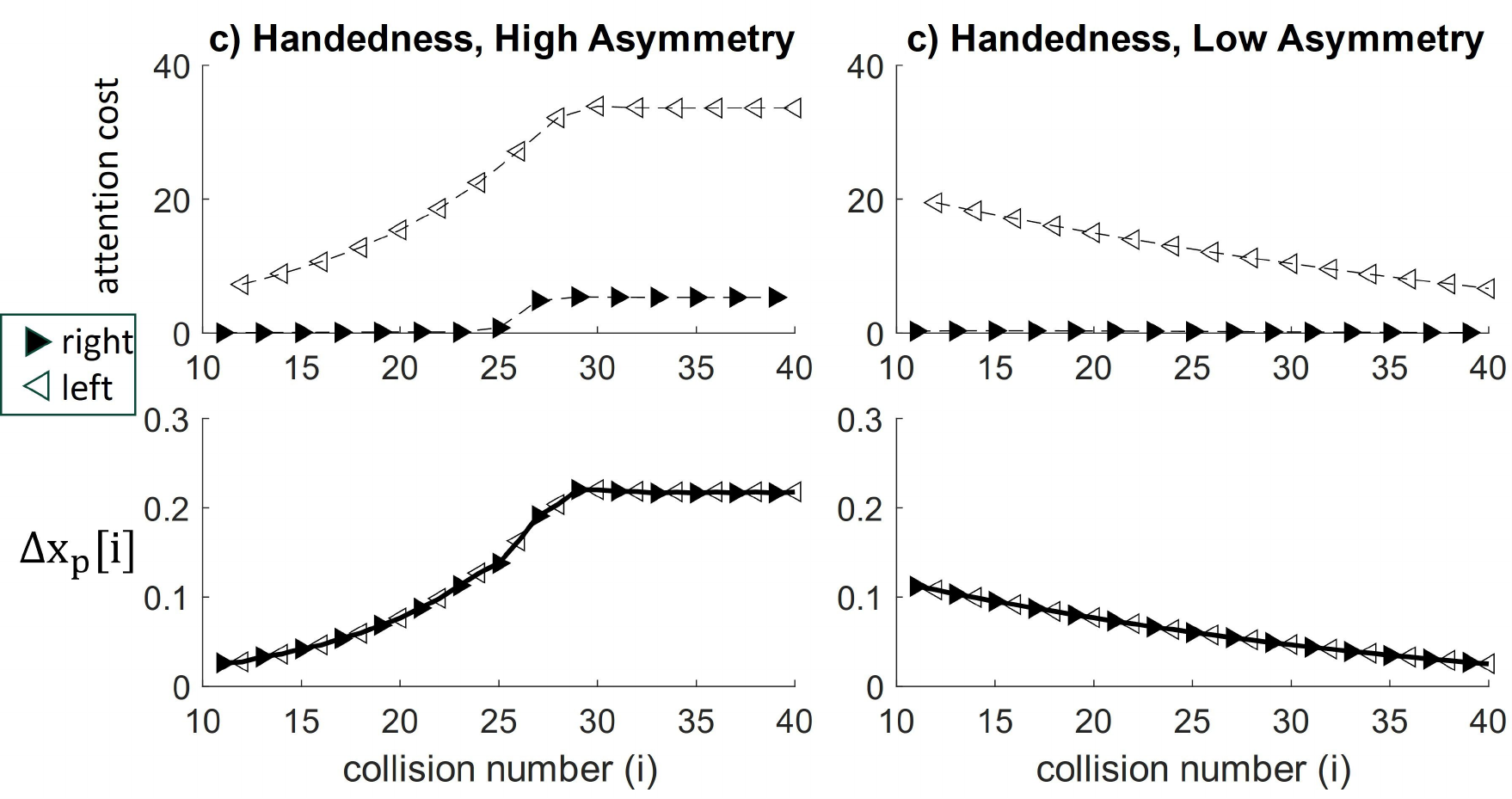}
\caption{Attention cost and net paddle movement for right-handedness, $i=11...40$. The damping factor $\delta$ difference between actuators for Low and High asymmetry is 1 and 3 respectively. In both cases, the left attention exceeds the right to compensate for its lack of mobility and accuracy. Attention and $\Delta x$ converges to 0 only in ``Low Asymmetry".}
\label{fig:hand_zoom}
\end{figure}

\begin{figure}
\includegraphics[trim = 0cm 0.7cm 0cm 1.4cm, width=\linewidth]{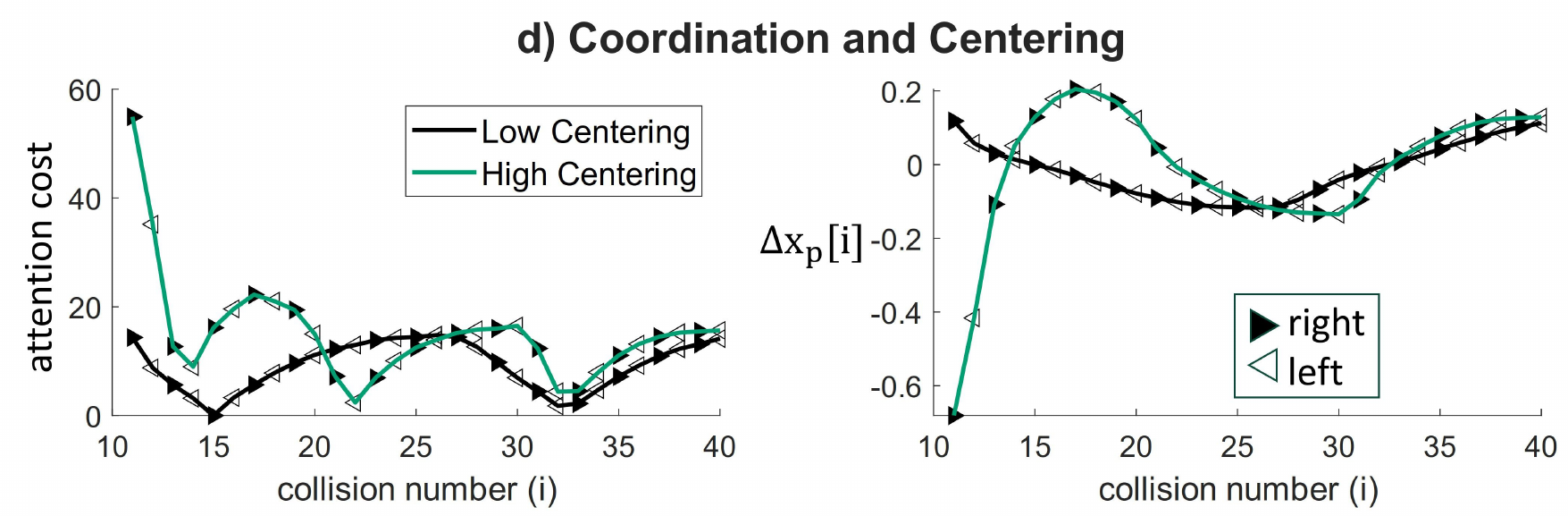}
\caption{\Red{Attention and net paddle movement for centering, $i=11...40$. We observe decaying oscillations in both plots, demonstrating a controller that can stabilize to the center. Higher frequency ripples occur in the ``High" case, where the centering/coordination tradeoff leans towards centering.}}
\label{fig:center_zoom}
\end{figure}

To successfully rally the ball between the two paddles as long as possible, the controller should stabilize the system towards ``Easy" tasks, as described in the previous section. As the tasks get easier with each collision, the attention cost should decrease. Ideally, the system is driven to a point where the paddles are not moving at all and the ball \Red{has no vertical velocity}. We examine \Red{emergent attention strategies} from solving (\ref{eqn:attention_problem_full}) at every collision in four scenarios. Fig.~\ref{fig:att_u_plots} plots the results of these four cases over 40 collisions ($i=1,...,40$), and animations corresponding to each case in Fig.~\ref{fig:att_u_plots} can be found at: \url{https://drive.google.com/drive/folders/1eXuQdB6YJW7x7cy4y_cFl9RHGLswZ1IP?usp=sharing}. All four cases start from the same initial conditions (ball velocity and paddle positions). In order to see the steady-state behavior of \textit{c)} and \textit{d)} more clearly, Fig.~\ref{fig:hand_zoom} and \ref{fig:center_zoom} plot the results for ($i=11,...,40$).

\begin{table}[!h]
\begin{center}
\caption{Sum of Costs over 12 collisions}
\label{tab:total_costs}
\begin{tabular}{ |c|cc| } 
   \hline

   & Attention (\textbf{q}$_p^\top$\textbf{q}$_p$) & Control ($r$\textbf{u}$^\top$\textbf{u})\\ 
  \hline
  a) No Coordination & 5225 & 878 \\ 
  \hline
  b) Weak Coordination & 718 & 39 \\ 
  \hline
  b) Strong Coordination & 515 & 35 \\ 
  \hline
  c) Low Asymmetry & 450 & 14\\ 
  \hline
  c) High Asymmetry & 682 & 39 \\ 
  \hline
  d) Low Centering & 1191 & 32 \\ 
  \hline
  d) High Centering & 1557 & 43 \\ 
  \hline
\end{tabular}
\end{center}
\end{table}

\paragraph{a) No Coordination:}\Red{$J_{vA}$ is set to 0, letting the paddle hit the ball with no consideration for the other paddle. This is equivalent to solving (\ref{eqn:position_problem}) at each collision.} The task difficulty increases at every collision, which eventually leads to an infeasible task (ball is too far away for the paddle). Since the paddle velocity at the point of collision is not being controlled, the paddle keeps adding energy every collision, constantly increasing the ball's vertical velocity. 

\paragraph{b) Coordination:} In the case where coordination is added \Red{($J_{vA}$ added to cost)}, the attention cost is able to decrease over time, but if the coordination is weak (coordination strength is adjusted with $\upepsilon$ in (\ref{eqn:qL_estimate})), the paddles and the ball will slowly drift away from the origin and not stabilize ($\Delta x_p[i]$ converges to nonzero value). This can be interpreted as a case where an attempt is made to coordinate, but if the agent does not have the ability to coordinate precisely enough, \Red{the ball still drifts away even if it slows down.} In both the strong and weak cases, adding coordination decreases both the overall attention cost and control effort (Table \ref{tab:total_costs}). This is expected since the coordination problem is able to extend its horizon over to the next collision, and therefore anticipate how difficult the problem is for the opposite paddle. 

\paragraph{c) Handedness:} To simulate right-handedness, we make the damping factor $\delta$ lower for the right paddle. Since the tracking task for the left hand is more difficult, more attention is allocated to the left side to compensate. Naturally, the attention differences between the left and right in Fig.~\ref{fig:att_u_plots} and \ref{fig:hand_zoom} are more distinct in the case where the asymmetry is high. In Fig.~\ref{fig:hand_zoom}, we also see that for low asymmetry the attention cost and net paddle movement converges towards 0, while the high asymmetry case can only converge to nonzero values. Since coordination is intuitively more difficult for the high asymmetry case, this behavior is expected. In Table \ref{tab:total_costs}, the low asymmetry case has the lowest control cost. This can be attributed to the particularly high agility of the right hand, which requires far less effort to control.

\paragraph{d) Centering:} \Red{Equation (\ref{eqn:v_des}) is the target velocity $v_d$ as a function of the target position $x_{pL}[N]$ (position of opposing paddle). If we multiply $x_{pL}[N]$ by a fraction, the target position can be adjusted between 0 and $x_{pL}[N]$, representing a tradeoff between coordination and aiming for the center. Not only does the agent have to minimize attention over multiple collisions, but the ball must head towards the 0 vertical position. This additional ``centering" goal ensures that the paddles stay within a reasonable range of motion and combats the drift seen in the weak coordination case. In Fig.~\ref{fig:att_u_plots} and \ref{fig:center_zoom}, the ``Low" and ``High" centering cases use 0.73 and 0.55 of $x_{pL}[N]$ in the cost function, respectively. Decaying oscillations emerge in both the attention and net paddle movement (Fig.~\ref{fig:center_zoom}), where oscillations are higher frequency in the ``High" case due to aiming for the center more aggressively. From Table \ref{tab:total_costs}, we also see that an aggressive strategy accrues higher costs. Both cases extend the system's settling time, allowing the ball position to descend. From Fig.~\ref{fig:att_strat}, we can see that the attention in cases \textit{b)} and \textit{c)} descends more quickly compared to \textit{d)}. The attention cost is inherently higher, not only because the task is more complex, but also because the settling time of the problem is extended. Out of all the cases, this one is the most robust to initial conditions.}

\section{Conclusions and Future Work}
We present a bio-inspired two-layered hierarchy of optimization problems that can rally a ball between paddles. The paddles can represent hands, arms, virtual objects, or real paddles. \Red{We find several emergent behaviors; stronger coordination leads to lower long-term attention, while high asymmetry and aggressive centering for stability increase overall attention. Our pong model has significant simplifications. The rail constraint allows the linear dynamics to be incorporated in the attention constraints; other nonlinear actuators that are common in sensorimotor models can use this framework if linearized. While the ball's dynamics are simple, other object-collision dynamics can fit into the existing framework, since the optimization relies on the ball trajectory regardless of how it is acquired.} 

Future work can expand the pong model to many other variations. Juggling can be simulated by increasing the number of objects to the model. Different objects can swapped in as well, such as a devil-stick \cite{devilstick_control} or slinky. A spectrum of coupling strengths between actuators can be explored, since actuators are not always completely independent. Attempting to further quantify and model the behavior observed from Fig.~\ref{fig:att_u_plots} could open up opportunities for more theoretical insights, so the stability of the system can be guaranteed.

\bibliography{citations}
\bibliographystyle{IEEEtran}

\end{document}